\documentclass[mathleft]{an}
\usepackage{amsmath}
\usepackage{amssymb}
\usepackage{textcomp}
\usepackage{graphicx}
\usepackage{times}
\usepackage{multirow}
\usepackage[flushleft]{threeparttable}

\begin{document}

\renewcommand{\figurename}{Fig.\,} 
\renewcommand{\tablename}{Tab.} 

\Pagespan{1}{}
\Yearpublication{2018}
\Yearsubmission{2017}

\DOI{10.1002/asna.201713383}

\title{The Radial and Rotational Velocity of $\zeta$\,Ophiuchi\thanks{Based on observations obtained with telescopes of the University Observatory Jena, which is operated by the Astrophysical Institute of the Friedrich-Schiller University.}}

\author{T.~Zehe\thanks{Corresponding author: \email{tamara.zehe@uni-jena.de}\newline}, M. Mugrauer, R. Neuh\"{a}user, A. Pannicke, O. Lux, R. Bischoff, D. W\"{o}ckel, D. Wagner}

\titlerunning{The Radial and Rotational Velocity of $\zeta$\,Ophiuchi}
\authorrunning{Zehe et al.}
\institute{Astrophysikalisches Institut und Universit\"{a}ts-Sternwarte Jena, Schillerg\"{a}{\ss}chen 2-3, 07745 Jena, Germany}

\received{8 June 2017}
\accepted{18 September 2017}
\publonline{later}


\abstract{We present the results of our spectroscopic monitoring campaign of $\zeta$\,Oph, carried out at the University observatory Jena with the spectrograph FLECHAS between July 2015 and September 2016. The main aim of the project is to determine the radial velocity of the star precisely, as a wide range of radial velocity measurements of $\zeta$\,Oph are given in the literature. Beside the proper motion and parallax, the radial velocity is an important input parameter for the calculation of the space motion of the star. $\zeta$\,Oph is considered to be a runaway star and  potential former companion of the progenitor star of the pulsar PSR\,J1932+1059, which both may have been located in a binary system and were then released during a supernova explosion about 1\,Myr ago in the Upper Scorpius OB association. For the calculation that the pulsar and $\zeta$\,Oph could have been at the same place, a radial velocity of $-9.0$\,km/s for $\zeta$\,Oph was used in the literature. By analysing the hydrogen and helium lines in 48 FLECHAS spectra, we determined the radial velocity of $\zeta$\,Oph to be $12.2\pm3.3$\,km/s and the projected rotational velocity to $432\pm16$\,km/s with no indication for variability.}

\maketitle

\section{Introduction}

For a sample of a few nearby neutron stars and a few OB-type runaway stars, whose proper motion and parallax as well as the radial velocity (RV) were well known, Hoogerwerf et al. (2000, 2001)\nocite{hoogerwerf2000,hoogerwerf2001} calculated their flight paths backwards in time to find cases where a runaway star and a neutron star could have been at the same place at the same time. This is possible because the proper motion of runaway stars, which is high compared to the velocity of the surrounding interstellar medium, often points exactly away from the star association from where they were ejected, either by a close encounter between two binary systems or a supernova explosion. The aforementioned cases would be evidence for a supernova in a binary, which formed a neutron star and released the former companion of the supernova progenitor star to become a runaway star. Since RVs of neutron stars are unknown in most cases (except, e.g. when the inclination of a bow shock is known, see e.g. Tetzlaff et al. 2011)\nocite{tetzlaff2011}, the result is probabilistic and true only for a certain RV of the neutron star (see Hoogerwerf et al. 2000, 2001\nocite{hoogerwerf2000,hoogerwerf2001}, for further details). The only case, where a runaway star and a neutron star could have been connected was $\zeta$\,Oph (alias 13\,Oph, HD\,149757, or HIP\,81377), with a distance of $112\pm3$\,pc ($\pi=8.91\pm0.20$\,mas, van Leeuwen 2007)\nocite{vanleeuwen2007}, the closest known O-type star, and the pulsar PSR\,J1932+1059, which may have been released in a supernova about 1\,Myr ago in the Upper Scorpius OB association, as described by Hoogerwerf et al. (2000, 2001)\nocite{hoogerwerf2000,hoogerwerf2001}. This worked well only when increasing the error bars for the measured (input) data, as they were known at that time (2000) to $10\,\sigma$ (see also Tetzlaff et al. 2011)\nocite{tetzlaff2011}. Later-on, when more precise values for proper motion and parallax of the pulsar became known, including smaller error bars, it did not work well anymore (Bobylev 2008; Tetzlaff et al. 2010; Kirsten et al. 2015)\nocite{bobylev2008, tetzlaff2010, kirsten2015}. In this context it is important to mention that a wide range of RVs (from $-35$\,km/s to $+30$\,km/s, as summarized in Tab.\,\ref{tab:RVVergleiche}) have been published for $\zeta$\,Oph in the literature. That includes individual RV measurements, but also average RVs derived from different sets of RV measurements, compiled by the authors. Hoogerwerf et al. (2000, 2001)\nocite{hoogerwerf2000,hoogerwerf2001} and Tetzlaff et al. (2010, 2011)\nocite{tetzlaff2010,tetzlaff2011} all used a RV of $-9.0\pm5.5$\,km/s for their analysis of $\zeta$\,Oph.

\begin{table}
\centering
\caption{Radial velocities of $\zeta$\,Oph from the literature with their uncertainties and Julian dates (JD) where available.}
\begin{tabular}{ccc}
\hline
RV $[$km/s$]$ & JD         & Reference\\
\hline
$-13.0\pm5.0$ & ---        & \cite{bobylev2006}\\
\\
$+15\pm4$     & 2447646    & \cite{reid1993}\\
\\
$-9.0\pm5.5$  & ---        & \cite{barbierbrossat1989}\\
\\
$+11\pm3$     & 2444326    & \cite{ebbets1981}\\
\\
$-12.6\pm4.1$ & 2440046    & \cite{conti1977}\\
\\
$-10\pm2.7$   & 2434944    & \multirow{3}{*}{$\left\}\begin{array}{c} \\ \\ \\ \end{array}\right.$\cite{buscombe1960}}\\
$-6\pm6.7$    & 2435296    & \\
$+12\pm8.5$   & 2435647    & \\
\\
$-19\pm5$     & ---        & \cite{wilson1953}\\
\\
$+4.6\pm8.6$  & 2442560    & \multirow{6}{*}{$\left\}\begin{array}{c} \\ \\ \\ \\ \\ \end{array}\right.$\cite{garmany1980}}\\
$+5.4\pm9.6$  & 2442561    & \\
$-17.5\pm10$  & 2442562    & \\
$-0.1\pm10.2$ & 2442566    & \\
$+2.3\pm14.3$ & 2442567    & \\
$-3.2\pm9.1$  & 2442568    & \\
\\
$+8.5$        & 2444771    & \cite{harmanec1989}\\
\\
$-20$         & ---        & \multirow{4}{*}{$\left\}\begin{array}{c} \\ \\ \\ \\ \end{array}\right.$\cite{abt1972}}\\
$-35$         & ---        & \\
$+30$         & ---        & \\
$-15$         & ---        & \\
\hline
\end{tabular}                                              	
\label{tab:RVVergleiche}
\end{table}

As $\zeta$\,Oph, which exhibits an apparent magnitude of $V=2.54$\,mag (\cite{perryman1997}) and a spectral type of O9.5V (Ebbets 1981 \& Marcolino et al. 2009)\nocite{ebbets1981}\nocite{marcolino2009}\footnote{Alternatively the star is classified as O9.5 giant by G\l{}\c{e}bocki \& Gnaci\'{n}ski (2005)\nocite{glebocki2005}, and  O9.2 sub giant by Sota et al. (2014)\nocite{sota2014}, respectively.}, is too bright for the GAIA satellite\footnote{In the future it might be possible to extract also some kinematic information of $\zeta$\,Oph from unsaturated GAIA pixels.} for normal data reduction, we started a spectroscopic monitoring campaign of the star to redetermine and constrain its RV. Once an accurate RV and later on possibly also GAIA data are available for $\zeta$\,Oph, the origin of $\zeta$\,Oph and the pulsar PSR\,J1932+1059 can be reinvestigated. Even if the pulsar progenitor star is not the former companion of $\zeta$\,Oph, it is possible that the pulsar and/or $\zeta$\,Oph originated from the Upper Scorpius OB association.

In this paper we present the results of our spectroscopic monitoring campaign of $\zeta$\,Oph, carried out at the University Observatory Jena. In section 2 we describe in detail the observations, followed by the data reduction and analysis in section 3. We summarise and discuss the results obtained in the course of our observing project in the last section of this paper.

\section{Observations}	

We initiated our spectroscopic monitoring campaign of $\zeta$\,Oph in July 2015, utilizing the fibre-linked \'Echelle spectrograph FLECHAS (\cite{mugrauer2014}), which has a resolving power of $R\approx9300$ and is operated at the Nasmyth focus of the 90\,cm telescope of the University Observatory Jena, which is located close to the small village Gro{\ss}schwabhausen, west of the city of Jena (Pfau 1984)\nocite{pfau1984}. The observatory was established at the Friedrich-Schiller University in 1962 and is utilized since that time for astronomical observations during all clear nights. In the course of our observing project, 48 spectroscopic observations of $\zeta$\,Oph were taken in 46 nights between July 2015 and September 2016. For all observing nights, spectra of the star were obtained with FLECHAS at the lowest possible airmasses, ranging between 2.1 and 2.9. The minimal possible airmass for $\zeta$\,Oph at the location of the University Observatory Jena is $X=2.1$.

In each observing epoch three spectra with an individual integration time of 150\,s were taken of $\zeta$\,Oph, in order to achieve a sufficiently high $S/N$ ratio and to optimally remove cosmics detected in the individual spectra. Using this observing setup, spectra with $S/N$ ratios in the range of 297 to 719 ($S/N\sim537$ at $\lambda=5836$\,\AA, on average) were taken with FLECHAS. Right before the spectroscopy of the star, for calibration purposes, always three well-exposed flat-field frames of a tungsten lamp were taken, as well as three spectra of a ThAr-lamp (with more than 700 lines detected) for the wavelength calibration, each with an integration time of 5\,s. In addition, for the dark current and bias subtraction, three dark frames with integration times of 5 and 150\,s were taken in each observing night during evening or morning twilight, respectively.

\section{Data Reduction and Analysis}

The reduction of the obtained spectroscopic data was performed with a software pipeline that was developed at the Astrophysical Institute Jena and is optimized for the reduction of data taken with FLECHAS. The pipeline is based on \verb"IRAF" (\cite{tody1986}), \verb"ESO-MIDAS" (\cite{banse1983}) and \verb"ESO-ECLIPSE" (\cite{devillard2001}) commands and performs the dark subtraction and flat-fielding of all spectroscopic data, as well as the extraction, wavelength calibration and averaging, including removal of cosmics, of all individual spectral orders (\cite{mugrauer2014}).

\begin{figure*}
\centering
\resizebox{\hsize}{!}{\includegraphics{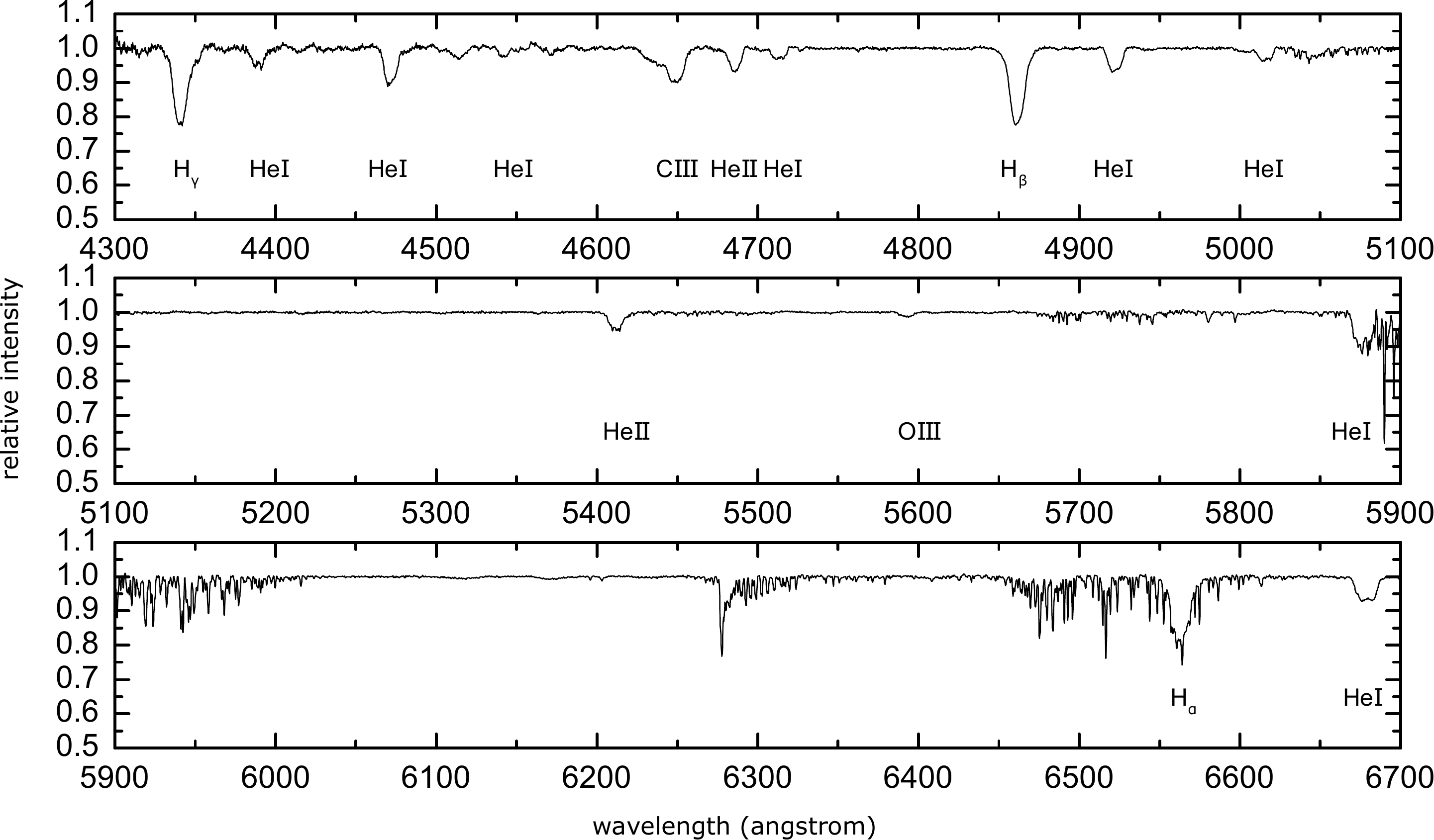}}
\caption{Normalized spectra of $\zeta$\,Oph taken with FLECHAS on 2016 June 23. Prominent absorption lines are indicated in the spectrum, among which are
the hydrogen and helium lines.}
\label{fig:SpektrumZetaOph}
\end{figure*}

Fig.\,\ref{fig:SpektrumZetaOph} shows exemplary the normalized and order-combined version of a fully reduced FLECHAS spectrum of $\zeta$\,Oph with the most prominent absorption lines indicated, among them the hydrogen and helium lines, whose reference wavelengths are also listed in Tab.\,\ref{tab:verwendeteLinien}.

\begin{table}
\caption{Reference wavelengths of neutral and ionized helium and hydrogen lines (used in this spectroscopic study), sorted by wavelength, with their equivalent width ($W_{\lambda}$), as detected in the FLECHAS spectra of $\zeta$\,Oph.}
\begin{tabular}{lccc}
\hline
Spectral Line      & $\lambda_0$ $[$\AA$]$ & $W_{\lambda}$ $[$\AA$]$ & Reference for $\lambda_0$\\
\hline
\ion{H}{$_\gamma$}  & 4340.48               &  2.64                   & \cite{lang1974} \\
\ion{He}{I(4387)}  & 4387.93               &  0.44                   & \cite{moore1945} \\
\ion{He}{I(4471)}  & 4471.48               &  0.89                   & \cite{moore1945} \\
\ion{He}{II(4541)} & 4541.59               &  0.19                   & \cite{moore1945} \\
\ion{He}{II(4686)} & 4685.68               &  0.55                   & \cite{moore1945} \\
\ion{He}{I(4713)}  & 4713.15               &  0.31                   & \cite{underhill1995} \\
\ion{H}{$_\beta$}   & 4861.34               &  2.52                   & \cite{lang1974} \\
\ion{He}{I(4921)}  & 4921.93               &  0.67                   & \cite{moore1945} \\
\ion{He}{I(5015)}  & 5015.68               &  0.34                   & \cite{moore1945} \\
\ion{He}{II(5412)} & 5411.52               &  0.52                   & \cite{moore1945} \\
\ion{He}{I(6678)}  & 6678.15               &  0.93                   & \cite{moore1945} \\
\hline                                                     	
\end{tabular}                                              		
\label{tab:verwendeteLinien}                               		
\end{table}

Alongside three Balmer lines (\ion{H}{$_\alpha$}, \ion{H}{$_\beta$} and \ion{H}{$_\gamma$}), 10 absorption lines of neutral and singly ionized helium are detected in the wavelength range covered by FLECHAS. Among these lines, the \ion{He}{I(4541)} line was not analysed due to its intrinsic weakness (equivalent width of only 0.19\,\AA\,in average). In addition the \ion{H}{$_\alpha$} line at $\lambda=6562.81$\,\AA\,and the \ion{He}{I} line at $\lambda=5875.62$\,\AA\,were also not analysed owing to telluric-line contamination. All spectral lines are significantly broadened, and most of them exhibit narrow emission and absorption peaks in their line profiles, which vary in strength and wavelength over time. These well-known line profile variations in the spectrum of $\zeta$\,Oph (\cite{walker1979}) are most probably caused by temperature variations in the stellar photosphere, induced by its nonradial pulsation.

For the analysis of the individual spectral lines standard \verb"IRAF" routines were used. Owing to the strong broadening of the individual spectral lines, their centroid wavelengths $\lambda_{cen}$ were determined by integrating the detected flux-levels $F_\lambda$ over the whole line profiles:

\begin{align}
\lambda_{cen} = \frac{\int \lambda F_{\lambda}~d\lambda}{\int F_{\lambda}~d\lambda}
\end{align}

\noindent The centroid wavelength of a spectral line together with its reference wavelength $\lambda_0$ yields the corresponding RV of the star and can be calculated via

\begin{align}
\text{RV} = c \cdot\frac{\lambda_{cen}-\lambda_0}{\lambda_0} + BC
\end{align}

\noindent were $c$ is the speed of light and $BC$ the barycentric correction, which rectifies for the RV offset between the observing site and the barycenter of the solar system at the observing epoch. The RVs obtained from the analysis of the individual spectral lines are summarized for all observing epochs in Tab.\,\ref{tab:RVErgebnisseLinien}, which also lists the average RVs and standard deviations for all lines. The RV standard deviations for all spectral lines are comparable to the typical uncertainties of their RV measurements, taken at all observing epochs.

\begin{table*}
\centering
\begin{threeparttable}
\caption{Radial velocities of $\zeta$\,Oph for all observing epochs obtained from the analysed hydrogen and helium lines in the FLECHAS spectra of the star. We have listed the modified barycentric Julian date ($BJD-2450000$) of the individual observing epochs, the airmass ($X$), and the $S/N$-ratio (at $\lambda=5836$\,\AA) of the spectra, as well as the determined radial velocities\tnote{3} of all lines in km/s. The average ($<$x$>$) and standard deviation ($\Delta$x) of all quantities are listed in the last two rows of the table.}
\begin{tabular}{ccccccccccccc}
\hline
Observing    & \multirow{2}{*}{$X$} & \multirow{2}{*}{$S/N$} & \multirow{2}{*}{\ion{H}{$_\gamma$}} & \ion{He}{I} & \ion{He}{I} & \ion{He}{II} & \ion{He}{I} & \multirow{2}{*}{\ion{H}{$_\beta$}} & \ion{He}{I} & \ion{He}{I} & \ion{He}{II} & \ion{He}{I}    \\
Epoch        &     &     &              & (4387)     & (4471)     & (4686)     & (4713)     &            & (4921)     & (5015)     & (5412)     & (6678) \\
\hline
BJD-2450000  &     &     & $[$km/s$]$   & $[$km/s$]$ & $[$km/s$]$ & $[$km/s$]$ & $[$km/s$]$ & $[$km/s$]$ & $[$km/s$]$ & $[$km/s$]$ & $[$km/s$]$ & $[$km/s$]$\\
\hline
7213.39      & 2.1 & 379 & 21.02        & 49.89      & -26.16     & 19.97      & 14.99      & ~0.70      & 13.90      & 24.61      & 13.02      & 23.90 \\
7214.39      & 2.1 & 573 & 24.30        & 40.18      & -10.95     & 20.30      & 23.88      & ~-1.65~    & 12.89      & 49.04      & 15.12      & 26.58 \\
7220.40      & 2.2 & 565 & 22.77        & 39.92      & -16.61     & 23.87      & ~4.54      & ~1.90      & ~6.79	  & 53.21      & 19.57      & 30.15 \\
7221.40      & 2.3 & 521 & 23.64        & 49.33      & ~-2.97     & ~9.20      & ~-1.36~    & ~1.21      & 15.30      & 59.02      & ~8.21      & 21.81 \\
7233.38      & 2.3 & 436 & 20.41        & 41.79      & -10.91     & ~7.11      & ~6.23      & ~-3.15~    & 15.84      & ~-0.11~    & ~4.77      & 12.96 \\
7500.55      & 2.2 & 699 & ~8.03        & 37.89      & -32.96     & 12.76      & -24.09~    & ~-9.26~    & ~3.20	  & 12.91      & 11.45      & ~1.15 \\
7507.56      & 2.1 & 646 & 15.45        & 79.59      & -16.74     & 15.44      & 27.94      & ~-1.22~    & ~8.24      & 18.36      & 15.59      & 18.95 \\
7511.50      & 2.2 & 662 & 18.58        & 93.15      & -22.73     & ~7.68      & 17.03      & ~3.28      & 19.76      & 26.28      & 15.10      & 23.44 \\
7516.54      & 2.1 & 591 & 19.27        & 45.71      & -14.95     & 11.89      & ~0.12      & ~-0.25~    & ~-1.22~    & 10.43      & 22.64      & 17.57 \\
7517.60      & 2.3 & 538 & 26.67        & 45.13      & ~-9.54     & ~9.18      & ~-0.21~    & ~-0.13~    & 13.01      & 31.94      & 24.87      & 25.26 \\
7519.48      & 2.2 & 524 & 23.84        & 44.40      & -13.27     & 14.43      & ~1.58~     & ~4.34      & ~6.71	  & 18.54      & 15.45      & 14.52 \\
7520.46      & 2.4 & 451 & 18.18        & 58.92      & -18.65     & 11.47      & ~-6.05~    & ~-2.88~    & ~9.80	  & 25.26      & 10.59      & ~9.82 \\
7527.49      & 2.1 & 432 & 27.30        & 61.81      & -16.43     & 17.93      & 11.30      & ~8.74      & 11.34      & 27.67      & 23.52      & 24.70 \\
7528.54      & 2.1 & 605 & 21.73        & 78.10      & -22.82     & 14.60      & 14.58      & ~1.26      & 16.78      & 49.54      & 15.74      & 23.44 \\
7535.45      & 2.2 & 502 & 16.96        & 66.48      & -26.37     & ~8.25      & 11.92      & ~-7.32~    & ~6.46	  & 22.91      & 25.67      & 11.94 \\
7546.51      & 2.2 & 509 & 25.16        & 39.68      & ~-7.23     & ~5.61      & 32.10      & ~0.41      & 26.18      & 27.42      & 10.23      & 25.71 \\
7547.39      & 2.4 & 517 & 19.70        & 39.12      & -16.60     & ~6.06      & ~7.36      & ~-7.17~    & 11.18      & 27.66      & 12.68      & 19.46 \\
7549.49      & 2.2 & 319 & 25.94        & 33.02      & -22.62     & 17.92      & ~-4.79~    & ~-1.14~    & ~5.02	  & ~6.75      & 19.39      & 20.62 \\
7555.49      & 2.3 & 503 & 23.76        & 44.99      & -31.77     & 18.91      & ~2.47      & ~0.24      & ~9.77	  & 17.89      & 24.44      & 22.19 \\
7562.40      & 2.1 & 703 & 23.27        & 68.02      & -17.17     & 11.70      & ~4.77      & ~-1.62~    & ~8.85	  & 39.03      & ~9.98      & 20.31 \\
7563.40      & 2.1 & 686 & 17.37        & 70.56      & -18.41     & ~6.12      & ~2.68      & ~-4.90~    & ~6.40	  & 27.56      & ~9.78      & 15.96 \\
7568.39      & 2.1 & 284 & 31.78        & 37.44      & -20.34     & 20.79      & ~-0.82~    & ~-1.53~    & ~5.25	  & 11.40      & 18.92      & 14.39 \\
7575.45      & 2.5 & 372 & 15.55        & 71.30      & ~-4.26     & ~9.91      & ~8.85      & ~-6.18~    & 10.92      & ~7.15      & 10.06      & ~6.11 \\
7576.44      & 2.3 & 443 & 23.14        & 63.83      & -11.71     & ~6.02      & ~-8.95~    & ~-3.37~    & 14.64      & 14.98      & ~8.52      & 20.52 \\
7588.37      & 2.1 & 684 & 17.38        & 42.75      & -26.79     & ~4.30      & -16.59~    & ~-1.60~    & 14.30      & 21.03      & ~1.67      & 19.18 \\
7589.37      & 2.1 & 651 & ~9.41        & 29.09      & -26.51     & ~7.88      & ~6.28      & ~-6.49~    & ~0.50	  & 15.91      & ~9.10      & 13.95 \\
7590.35      & 2.1 & 413 & 21.11        & 61.85      & -22.05     & ~1.44      & ~-6.90~    & ~-8.52~    & ~8.23	  & ~1.56      & ~3.07      & 13.90 \\
7591.36      & 2.1 & 297 & 18.38        & 63.56      & -19.01     & 14.13      & -25.05~    & ~-5.36~    & 11.71      & 14.38      & 23.38      & 22.71 \\
7597.36      & 2.2 & 394 & 16.90        & 56.19      & -19.41     & 12.62      & ~-1.34~    & ~-6.87~    & 12.42      & 27.41      & ~9.81      & 14.53 \\
7597.37      & 2.3 & 409 & 22.44        & 32.85      & -13.52     & 19.52      & -15.16~    & ~-2.29~    & ~2.56	  & 23.93      & ~6.51      & 16.22 \\
7611.33      & 2.3 & 437 & 21.86        & ~7.04      & -23.21     & ~5.24      & ~1.21      & ~-4.29~    & ~5.63	  & ~1.30      & ~9.78      & ~1.75 \\
7617.31      & 2.3 & 500 & 21.25        & 49.28      & -16.80     & ~5.85      & -11.74~    & ~-5.28~    & ~4.64	  & 23.74      & 13.34      & 10.02 \\
7618.31      & 2.3 & 595 & 24.06        & 84.28      & -15.97     & 12.31      & ~0.57      & ~-1.48~    & ~0.04	  & ~4.44      & 14.77      & 21.50 \\
7619.31      & 2.3 & 613 & 18.54        & 44.56      & -24.28     & ~9.95      & ~-0.57~    & ~-9.28~    & ~5.47	  & 19.07      & 18.13      & ~7.99 \\
7619.33      & 2.5 & 629 & 23.11        & 31.66      & -18.17     & ~8.16      & 12.99      & ~0.22      & 12.01      & ~5.20      & 12.05      & 24.27 \\
7624.31      & 2.4 & 681 & 14.19        & 63.24      & -25.06     & 15.56      & -23.75~    & ~-3.76~    & ~-5.36~    & 25.18      & 11.89      & ~8.50 \\
7625.31      & 2.4 & 719 & 23.18        & 44.41      & -21.22     & ~9.28      & ~-8.75~    & ~2.28      & 13.51      & 30.68      & ~6.44      & 25.49 \\
7626.30      & 2.4 & 710 & 22.55        & 53.31      & -19.35     & ~9.54      & 42.17      & ~-1.14~    & 18.97      & 21.63      & ~7.85      & 25.09 \\
7631.29      & 2.4 & 507 & 19.16        & 39.06      & -11.20     & ~8.75      & 16.36      & ~-7.11~    & ~9.71	  & 13.14      & 11.56      & 17.14 \\
7632.30      & 2.5 & 706 & 20.24        & 31.00      & -23.32     & 16.67      & 16.09      & ~-2.23~    & 15.15      & 35.63      & ~9.13      & 19.73 \\
7633.32      & 2.9 & 386 & 23.47        & 71.25      & -33.76     &-13.49~     & -17.14~    & -12.04~    & -11.68~    & 30.03      & 25.58      & ~6.11 \\
7638.28      & 2.5 & 559 & 31.51        & 42.75      & -10.01     & 10.21      & 13.53      & ~-0.28~    & 15.64      & 12.32      & 17.48      & 30.68 \\
7639.28      & 2.6 & 613 & 19.14        & 53.64      & -29.91     &~-1.04~     & -15.42~    & ~-8.06~    & ~5.81	  & 10.44      & ~6.40      & 18.56 \\
7640.27      & 2.5 & 675 & 14.93        & 44.61      & -26.85     & ~3.93      & ~-7.68~    & ~-6.38~    & ~-3.90~    & ~8.02      & ~3.47      & ~9.02 \\
7644.28      & 2.7 & 564 & 14.94        & 78.47      & -15.74     & ~6.07      & ~7.45      & ~-7.30~    & ~7.28	  & 24.22      & ~6.70      & ~9.08 \\
7645.26      & 2.5 & 567 & 18.48        & 35.56      & -18.90     & ~7.78      & ~-3.52~    & ~-7.02~    & ~6.10	  & 26.32      & ~6.55      & 14.48 \\
7646.27      & 2.7 & 509 & ~9.11        & 34.65      & -17.09     & -5.15      & -34.61~    & -13.12~    & ~-7.18~    & 21.08      & ~6.42      & ~5.42 \\
7653.25      & 2.7 & 509 & 15.47        & 31.60      & -20.10     & 14.20      & ~-9.17~    & ~-8.52~    & ~4.90	  & 15.26      & ~6.98      & 13.72 \\
\hline
$<$x$>$      & 2.3 & 537 & 20.3         & 50.6       & -19.0~     & 10.2       & ~1.4       & ~-3.2~     & ~8.4       & 21.7       & 12.8       & 17.1 \\
$\Delta$x    & 0.2 & 118 & ~5.0         & 17.2       & ~7.0       & ~6.9       & 15.2       & ~4.4       & ~7.2       & 13.2       & ~6.4       & ~7.3 \\
\hline
\end{tabular}
\begin{tablenotes}
\item[3] \footnotesize The typical RV uncertainties of individual line measurements are: 14.4, 6.6, 6.8, 5.0, 5.3, 3.2, 5.2, 10.3, 6.4, and 7.2\,km/s, given in the same order as in the table.
\end{tablenotes}
\end{threeparttable}

\label{tab:RVErgebnisseLinien}
\end{table*}

In addition to the centroid wavelengths we also measured the widths of the spectral lines in all FLECHAS spectra of $\zeta$\,Oph by determining the wavelengths at which the line profiles start to breach out of the noise floor at the continuum flux level. The measured line-width $\text{LW}$ of a spectral line yields the projected rotational velocity $vsin(i)$ of the star:

\begin{align}
vsin(i) =c \cdot \frac{\text{LW}/2}{\lambda_0}
\end{align}

\noindent assuming only rotation as the main line broadening mechanism. The resulting average LWs and estimates of the rotational velocities of $\zeta$\,Oph for all observing epochs obtained from the analysed spectral lines are summarized in Tab.\,\ref{tab:vsiniErgebnisseLinien},
together with their uncertainties.

\section{Results and Discussion}

In the course of our spectroscopic monitoring campaign of $\zeta$\,Oph between July 2015 and September 2016, well-exposed spectroscopic data ($S/N\sim 537$, on average) were collected with FLECHAS in 48 observing epochs.

\begin{figure}[h]
\centering
\resizebox{\hsize}{!}{\includegraphics{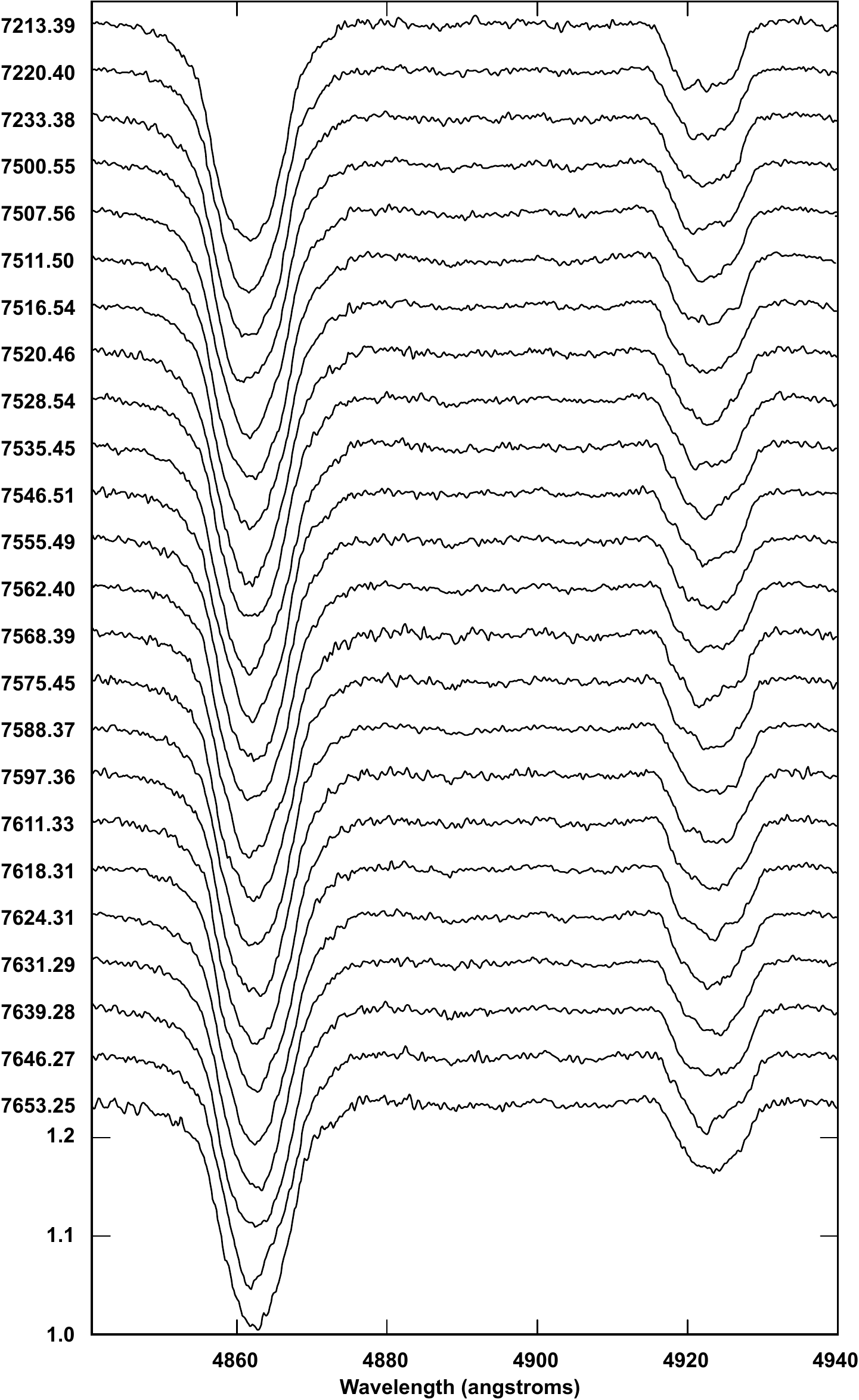}}
\caption{Wavelength range between $\sim 4840$ and 4940\,\AA\,of a selection of 24 spectra taken in the course of our observing campaign of $\zeta$\,Oph. The date of each observing epoch is given in the left-most position in the format $BJD-2450000$. The spectra are shifted along the y-axis with the corresponding scaling, which is shown in the lower left corner of the diagram. The line profile variations of the \ion{H}{$_\beta$}- and \ion{He}{I(4921)} lines are clearly visible between the individual observing epochs, while no clear line splitting, typical for a SB2 binarity, could be detected.}
\label{fig:sb2}
\end{figure}

While Cazorla et al. (2017)\nocite{cazorla2017}, Mason et al. (2009)\nocite{mason2009}, and Bobylev et al. (2006)\nocite{bobylev2006} classify $\zeta$\,Oph as a single star, Chini et al. (2012)\nocite{chini2012} report the star to be a double-lined (SB2) spectroscopic binary. However, as it is illustrated in Fig.\,\ref{fig:sb2} our spectroscopic data of the star does not show any evidence for binarity. Other than the known line profile variations of the star in particular no line splitting, which is expected for a SB2 binary, was detected throughout our observing campaign of $\zeta$\,Oph.

\begin{figure}
\centering
\resizebox{\hsize}{!}{\includegraphics{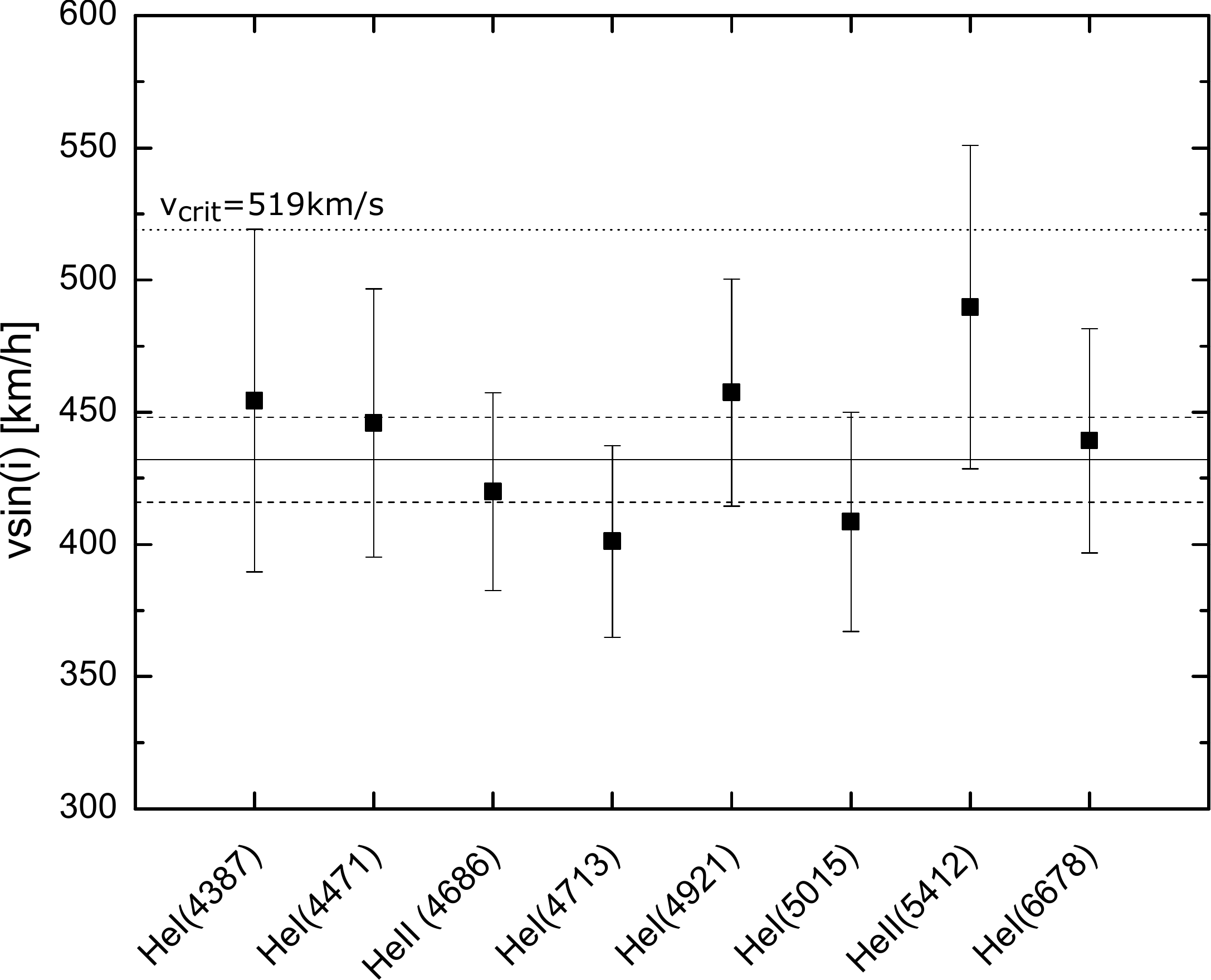}}
\caption{Average projected rotational velocities $vsin(i)$ with their uncertainties derived from the widths of the analyzed helium lines in the FLECHAS spectra of $\zeta$\,Oph. The projected critical rotational velocity $v_{\text{crit}}=519$\,km/s is illustrated as a dotted line. The solid line shows the weighted average of the projected rotational velocities derived from the widths of the helium lines, with its uncertainty indicated with dashed lines.}
\label{fig:vsini}
\end{figure}

We used the widths and centroid wavelengths of 10 hydrogen and helium lines, all well detected in the FLECHAS spectra between 4300 and 6700\,\AA, to derive the projected rotational velocity and the RV of the star. The average projected rotational velocities $vsin(i)$ of the star with their uncertainties, calculated from the widths of the analysed spectral lines, are illustrated in Fig.\,\ref{fig:vsini}. The hydrogen lines cannot be used for the determination of the rotational velocity because beside rotation broadening, these lines are also significantly Doppler-broadened. The projected rotational velocities, obtained from the widths of the helium lines all range between 400 and 490\,km/s.

\begin{table}
\centering
\caption{Determined average linewidth (LW) and derived rotational velocities $vsin(i)$ of $\zeta$\,Oph with their uncertainties.}
\begin{tabular}{ccc}
\hline
Spectral Line         & $\text{LW}$ $[$\AA$]$ & $vsin(i)$ $[$km/s$]$\\
\hline
\ion{He}{I(4387)}     & $13.3\pm1.9$          & $454.4 \pm 64.9$\\
\ion{He}{I(4471)}     & $13.3\pm1.5$          & $445.9 \pm 50.8$\\
\ion{He}{II(4686)}    & $13.1\pm1.2$          & $419.9 \pm 37.5$\\
\ion{He}{I(4713)}     & $12.6\pm1.1$          & $401.1 \pm 36.2$\\
\ion{He}{I(4921)}     & $15.0\pm1.4$          & $457.5 \pm 42.9$\\
\ion{He}{I(5015)}     & $13.7\pm1.4$          & $408.5 \pm 41.5$\\
\ion{He}{II(5412)}    & $17.7\pm2.2$          & $489.8 \pm 61.2$\\
\ion{He}{I(6678)}     & $19.6\pm1.9$          & $439.2 \pm 42.5$\\
\hline
\end{tabular}
\label{tab:vsiniErgebnisseLinien}
\end{table}

The break-up or critical rotational velocity of a star depends on its mass $M$ and radius $R$:

\begin{align}
v_{crit} = \sqrt{\frac{GM}{R}}
\end{align}

\noindent with the gravitational constant $G=6.67\cdot10^{-11}$\,m$^3$kg$^{-1}$s$^{-2}$ and with $M=13^{+10}_{-6}\,\text{M}_{\odot}$, $R=9.2^{+1.7}_{-1.4}\,\text{R}_{\odot}$, as estimated by  Marcolino et al. (2009)\nocite{marcolino2009} for $\zeta$\,Oph we expect a break-up velocity of about $v_{crit}=519^{+248}_{-168}$\,km/s. In addition, we estimated the uncertainty of $v_{crit}$ that originates from the different spectral types and luminosity classes of $\zeta$\,Oph given in the literature, using the typical stellar properties of late O- and early B-type dwarfs and giants, taken from \cite{cox2000}. We obtain $v_{crit}$ in the range between $\sim700$\,km/s for dwarfs and $\sim500$\,km/s for giants, which correspond well with the uncertainty of $v_{crit}$ derived above.

The projected rotational velocities derived from the widths of the \ion{He}{I} and \ion{He}{II} lines are all consistent with each other within their uncertainties; hence the weighted average of the projected rotational velocities is $vsin(i)=432\pm16$\,km/s. Therefore the star rotates near its break-up velocity and the minimal possible inclination of the stellar rotation axis is $i=56^{\circ}$ (respectively as low as $i=34^{\circ}$ considering the 1$\sigma$ uncertainty of $v_{crit}$). The obtained value of $vsin(i)$ is in good agreement with those obtained by others earlier, for example, $420\pm8$\,km/s by Harmanec (1989)\nocite{harmanec1989}, $\sim415$\,km/s by Walker (1991)\nocite{walker1991}, or $400\pm20$\,km/s by Reid et al. (1993)\nocite{reid1993}, while it deviates at the 2$\sigma$ level from $vsin(i)$ values reported by Cazorla et al. (2017)\nocite{cazorla2017} ($378\pm15$\,km/s), van Belle (2012)\nocite{vanBelle2012} ($\sim 385$\,km/s), and Howarth et al. (1997)\nocite{howarth1997} ($372\pm30$\,km/s). Zorec et al. (2016)\nocite{zorec2016}, and Sim\'{o}n-D\'{\i}az \& Herrero (2014)\nocite{SimonDiaz2014} obtained even smaller values of $344\pm34$\,km/s and $\sim319$\,km/s, respectively.

\begin{figure}[h]
\centering
\resizebox{\hsize}{!}{\includegraphics{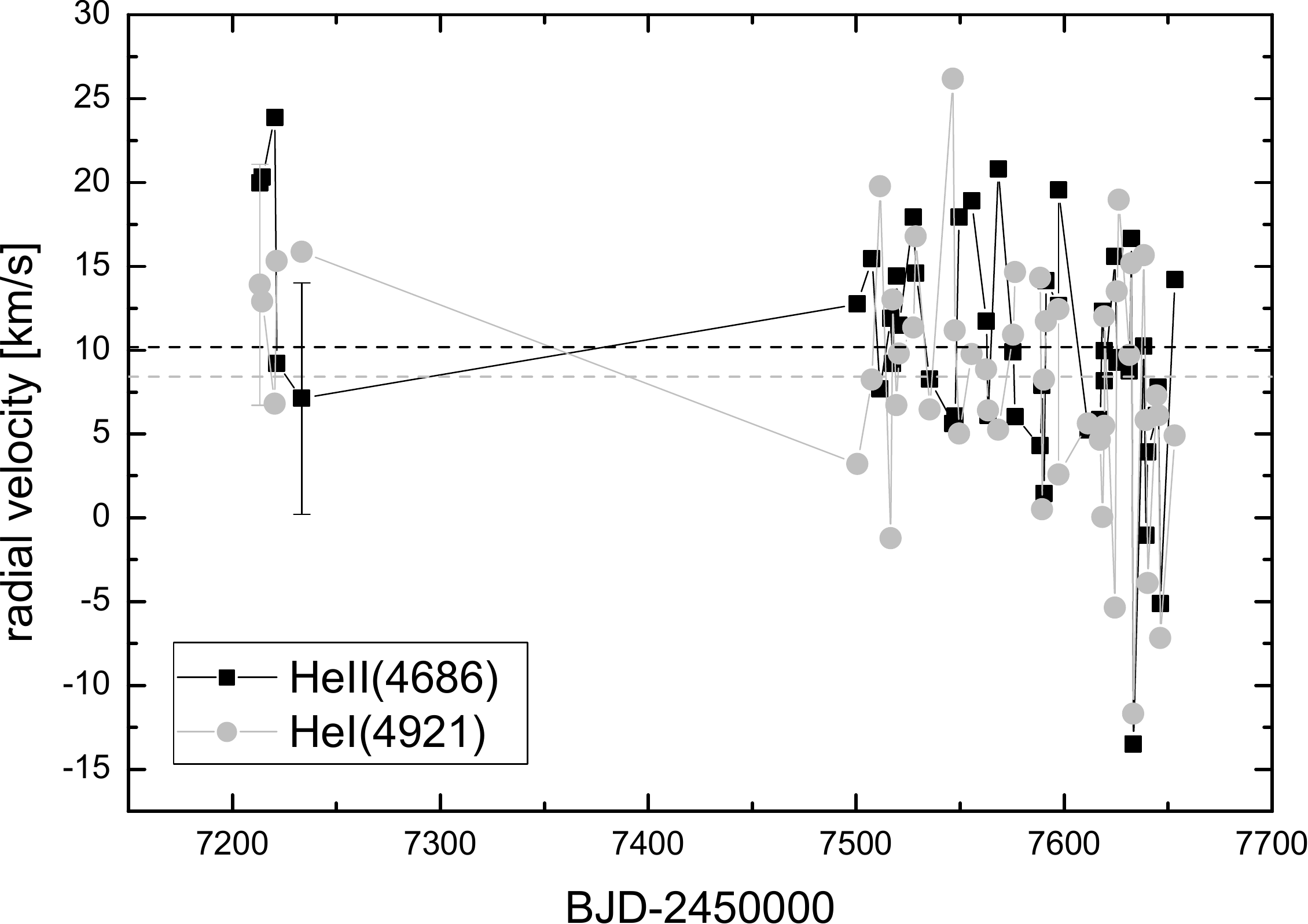}}
\caption{Radial velocities of $\zeta$\,Oph obtained from the analysis of the \ion{He}{II(4686)} (grey dots) and \ion{He}{I(4921)} line (black squares), as well as their average RV's (dashed lines) for all observing epochs. The typical uncertainty of the individual radial velocity measurements is about 7\,km/s. No significant trends are present in the radial velocities of the individual spectral lines.}
\label{fig:RVUnterschiede}
\end{figure}

In addition to their use for the determination of the rotational velocity of $\zeta$\,Oph the hydrogen and helium lines were also used to derive the RV of the star. As illustrated in Fig.\,\ref{fig:RVUnterschiede} with the \ion{He}{II(4686)} and \ion{He}{I(4921)} line no significant trends or variations are present in the RVs obtained from the analysis of the individual spectral lines throughout our spectroscopic monitoring campaign of the star. However, the RVs exhibit standard deviations of up to 17.2\,km/s, which is quite comparable to the typical uncertainties of the RV measurements obtained for the different spectral lines in all observing epochs.

Furthermore, the standard deviation of all obtained RV measurements is 19.9\,km/s, which is in the same order of magnitude as that of the published RVs of $\zeta$\,Oph (15 km/s), that is, the wide scatter of RV measurements found for the star in the literature could well be attributed to the RV fluctuations induced by the intrinsic width of the spectral lines of the star and/or their known profile variations.

\begin{figure}
\centering
\resizebox{\hsize}{!}{\includegraphics{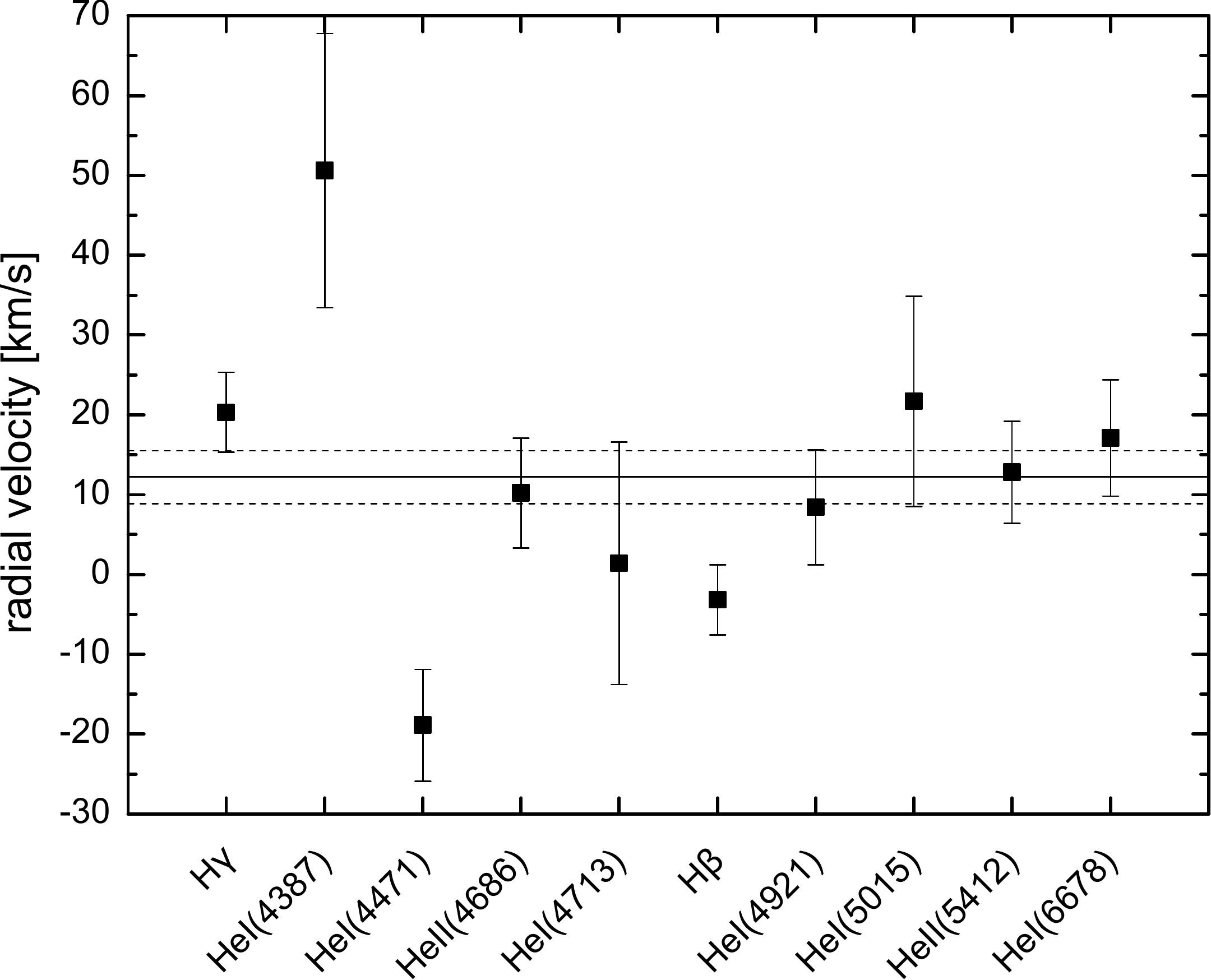}}
\caption{Radial velocities determined from the analysis of the hydrogen and helium lines, together with their weighted average (solid line) and uncertainty (dashed lines), excluding the \ion{He}{I(4387)} and \ion{He}{I(4471)} line.}
\label{fig:rv}
\end{figure}

In Fig.\,\ref{fig:rv} the RVs of the individual H and He lines are shown, together with their uncertainties. It is noticeable that the radial velocities of \ion{He}{I(4387)} and \ion{He}{I(4471)} deviate from the other determined RVs. However, this deviation is not due to the wavelength calibration of the instrument, which shows in all spectra over the whole covered spectral range a root mean square of only 20\,m\AA~(1.4\,km/s of absolute RV precision). In addition, the long-term RV stability of the instrument had already been established earlier, for example, by Irrgang et al. (2016)\nocite{irrgang2016} and recently by Bischoff et al. (2017)\nocite{bischoff2017}.
The average RV of all lines is $12$\,km/s and the RVs of most of the helium lines agree well within their uncertainties with the average RV value, that is, they show deviations from the average RV of less than 1$\sigma$, which is consistent with Gaussian noise. In contrast, the RVs of the H lines (\ion{H}{$_\beta$} and \ion{H}{$_\gamma$}) as well as the \ion{He}{I(4387)} and \ion{He}{I(4471)} line deviate from the average RV by more than 1.7$\sigma$, which is unexpected for the given number of RV measurements, assuming Gaussian noise only. For the RV determination, we therefore exclude these lines and obtain a weighted average of $12.2\pm3.3$\,km/s for the RV of $\zeta$\,Oph. This result is in good agreement with some RV determinations obtained earlier in other spectroscopic monitoring campaigns of the star, for example, $15\pm4$\,km/s by Reid et al. (1993)\nocite{reid1993}, or $11\pm3$\,km/s by Ebbets (1981)\nocite{ebbets1981}. Some of the negative RV values in the literature might have been obtained from the interstellar sodium lines (\ion{Na}{D1} line at $\lambda= 5895.94$\,\AA\,and the \ion{Na}{D2} line at $\lambda=5889.97$\,\AA\,) as we obtain $RV=-17.0\pm3.1$\,km/s for those.

With one of the previously published radial velocities of $\zeta$\,Oph ($-9.0 \pm 5.5$\,km/s, possibly coming from the interstellar Na lines), it was suggested that this star and the pulsar PSR\,J1932+1059 were located (about 1 Myr ago) at the same time and the same place inside the Upper Scorpius OB association, which would work for a certain radial velocity of the pulsar that is otherwise unknown. With Tetzlaff et al. (2010, 2011)\nocite{tetzlaff2010,tetzlaff2011} and Kirsten et al. (2015)\nocite{kirsten2015} doubting the common origin of these two stars, and with the positive radial velocity of $\zeta$\,Oph, it is even more clear that it is not a former companion of the PSR\,J1932+1059 progenitor released by its supernova. Their origins are unclear.

\begin{acknowledgement}
We would like to thank all observers who have participated in some of the observations of this project, obtained at the University Observatory Jena: in particular T. Heyne, S. Buder, H. Gilbert and C. G\"{u}ngor. We also acknowledge Kerstin Weis, Dominik Bomans and Matthias Ammler-von Eiff for their valuable comments on an earlier version of this manuscript. This publication makes use of data products of the \texttt{SIMBAD} and \texttt{VizieR} databases, operated at CDS, Strasbourg, France.
\end{acknowledgement}

\end{document}